\documentstyle[12pt,iopconf]{article}
\begin{document}
\title{Far-infrared and capacitance spectroscopy of self-assembled InAs quantum dots}

\author{A Lorke\dag\footnote{E-mail: axel.lorke@physik.uni-muenchen.de.}, 
M Fricke\dag, B T Miller\dag, M Haslinger\dag, J P Kotthaus\dag, G Medeiros-Ribeiro\ddag\ and P M Petroff\ddag}

\affil{\dag\ Sektion Physik, LMU M\"unchen, Geschwister-Scholl-Platz 1, \\ 80539 M\"unchen, GERMANY}

\affil{\ddag\ Materials Department and QUEST, University of California, \\ Santa Barbara, CA 93106, USA}

\beginabstract
We review a number of recent experiments which are probing ground state and excitations of few-electron systems in self-assembled InAs quantum dots. Far-infrared spectroscopy, together with local as well as large-scale capacitive probing allows for a detailed investigation of the different contributions to the many-particle spectrum in the dots. The influence of electron--electron interactions on the ground state and the excitations is discussed.
\endabstract

\section{Introduction}
With the development of the Stranski-Krastanow growth mode in recent years\cite{Goldstein,Leonard,Marzin,Drexler,Bimberg,Madhukar}, a powerful technique has emerged for the fabrication and experimental study of nm-size semiconductor systems. In these systems, the charge carriers are strongly confined in all three dimensions, so that they exhibit true "zero-dimensional" or "atomic" behavior. Contrary to real atoms, however, they are confined to a nearly parabolic potential with a characteristic length that is comparable to the magnetic length for magnetic fields accessible in the laboratory. Self-assembled Stranski-Krastanow quantum dots thus provide for a model system in which the interplay between compositional confinement, Coulomb interaction, and magnetic confinement can be studied in detail. Furthermore, through design, external bias or excitation strength, the number of carriers per dot can finely and {\it in situ} be tuned. A number of experimental studies have been carried out \cite{Marzin,Drexler,Bimberg,Rinaldi} to investigate the electronic properties of self-assembled dots by interband  and intraband spectroscopy as well as DC transport. 

Here we summarize recent experiments on both large and small scale dot arrays which probe the dots' many particle ground state and excitations by capacitance and far-infrared spectroscopy\cite{Fricke1,Fricke2,Barbara}. From the combined results of these complimentary techniques we derive a detailed picture of the contributions of compositional confinement, electron--electron interaction, and magnetic confinement to the many particle ground states and excitations.

\section{Experimental}
The samples are grown by molecular beam epitaxy. The InAs dots are embedded in a field-effect-transistor (FET) structure, which allows for an {\it in situ} tuning of the number of electrons per dot by  application of a suitable gate voltage\cite{Drexler,Medeiros}. Details of the procedure used for the formation of the InAs dots and the resulting structural properties can be found e.g. in Refs.\ \cite{Leonard,Medeiros}. We estimate the dots' diameter and height to be 20 nm and 7 nm, respectively, and the dot density to be $\approx 10^{10}cm^{-2}$. 

All experiments are carried out at liquid He temperatures with magnetic fields up to $B=15$ T applied perpendicular as well as parallel to the sample surface. The far-infrared response of the dots is measured in the energy range between 5 and 100 meV using a rapid-scan Fourier transform spectrometer. Large scale capacitance spectroscopy is carried out using standard lock-in technology. For small-scale capacitance probing a bridge technique, as described, e.g. in Refs.\ \cite{Schmerek} is employed. 
 
\section{Results and discussion}

\begin{figure}
\vspace{5.5 cm}
\caption{(a) Capacitance--voltage trace of a large-area sample ($A = 3.5$ mm$^2$) at $B=12$ T. (b)--(d) Normalized transmission at $B=12$ T, for $n=2$, 4, 6, in the energy range of the $\omega_+$ resonance. Note the qualitative change in the spectra as the $p$-state is filling.}
\label{FigKap+transm}
\end{figure}

Figure \ref{FigKap+transm}(a) displays typical capacitance data of a large-area sample at high magnetic fields, here $B_\perp = 12$ T. At very low gate bias, $V_g \leq -1$ V, the signal is given by the geometric capacitance between top gate and the doped GaAs back contact. At  $V_g \approx -0.9$\ V a sharp increase of the capacitance indicates the charging of the lowest electron state in the dots. Even though this state is doubly degenerate, the second electron is only loaded at a slightly higher gate voltage of $\approx -0.7$ V, because the 2-electron ground state is affected by the repulsive electron--electron interaction (here, we neglect spin-splitting, which is only a minute correction to the electronic states, see below). 

To model the electronic charging of our dots we use the following expression for the total energy of a dot, charged with $n$ electrons, embedded in a capacitor structure.

\begin{equation}
W_{n} =  \frac{Q^2}{2 C} + E_n - \frac{1}{4 \pi \varepsilon \varepsilon_0}  \frac{(n e)^2}{4 t_b} - \frac{n e \sigma_n}{\varepsilon \varepsilon_0} t_b 
\label{eq1}\end{equation}

The first term accounts for the energy of the charged capacitor (charge $Q$, capacitance $C$, surface charge density $\sigma$), the second for the $n$-electron ground state energy of the dot. The third and fourth term describe the influence of the dot's image charge and of the electric field inside the capacitor, respectively. 

It is easy to show from Eq. \ref{eq1} that the electron--electron interaction energy that leads to the double peak structure in the capacitance around $-0.8$ V is given by 

\begin{equation}
E_{1,2}=e\frac{t_b}{t_{tot}}\Delta V_g + \frac{e^2}{8\pi \varepsilon \varepsilon_0 t_b}
\label{eq2}\end{equation}

Here, $\Delta V_g$ is the difference in gate voltage between charging of the first and the second electron, $t_b$ is the distance between the back contact and the dots and $t_{tot}$ is the distance between the back contact and the gate. Because $t_{tot} \gg t_b$, we have neglected the image charge of the top gate. For the present sample with $t_b = 25$ nm and $t_b / t_{tot} = 1/7$, we find $E_{1,2} = 23.3$ meV. 

\begin{figure}
\vspace{5.5 cm}
\caption{Capacitance--voltage traces of a small-scale sample (area $A = 89\ \mu$m$^2$). The magnetic field was changed between $B = 0$ and 13 T in steps of $\Delta B = 1$ T. Curves are offset for clarity. Lines are indicating the dispersion of the $p$-states and the lowest $d$-state.}
\label{FigBarbara}
\end{figure}

In the present dots, which are oblate and nearly circular, the next higher (empty) state is expected to be fourfold degenerate at $B=0$. Because of its symmetry properties it is often called ''$p$-state'' (the lowest state accordingly being labeled $s$-state). In large-area samples, a lateral variation of the growth rate leads to a broadening which makes it impossible to distinguish individual peaks of the $p$-state. Reducing the probed area to below 100 $\mu$m$^2$, however, allows us to identify the individual $p$-state charging peaks and even the lowest $d$-state (Fig.\ \ref{FigBarbara})\cite{Barbara}. In such small-scale samples we also observe a rich substructure of smaller peaks which seem to be replicas of the main peaks indicated by broken lines in Fig.\ \ref{FigBarbara}. The origin of these replicas is not completely understood; monolayer fluctuations or dot--dot interactions are possible explanations \cite{Barbara}.  

Figure \ref{FigBarbara} shows the characteristic splitting of the $p$-levels when a strong magnetic field is applied perpendicular to the plane of the dots. This splitting is roughly linear in $B$ and we can extract an effective mass of $m^* = 0.063$ $m_e$. This value is considerably higher than the conduction band edge mass of InAs and can be attributed in part to the high non-parabolicity of this material and in part to the fact that a large fraction of the electron wave function leaks into the GaAs barrier layers \cite{Peeters}. 

\begin{figure}
\vspace{5.5 cm}
\caption{Far-infrared resonance positions for dots filled with two (a) and three (b) electrons. The partly filled $p$-state in (b) gives rise to additional $\omega_+$ transitions. Insets display possible resonances in a simple, single-particle picture.}
\label{FigFIRpos}
\end{figure}

Figure \ref{FigFIRpos} gives an overview of the measured far-infrared resonance positions as a function of magnetic field for different gate voltages. On large scale samples the far-infrared transmission and the capacitance can be recorded {\em simultaneously}, and we can through capacitance spectroscopy directly translate the bias voltages into electron occupation numbers, as shown in Fig.\ \ref{FigKap+transm}. For $n_e \approx 2$ (Fig.\ \ref{FigFIRpos}(a)) the characteristic two-mode spectrum of a parabolically confined electron system in a magnetic field is observed (cf. solid lines) \cite{minute_splitting}. No significant difference is found between $n_e = 1$ and $n_e = 2$, which further supports the assumption of an almost parabolic confinement, so that the generalized Kohn theorem applies \cite{Yip}. This situation changes drastically, when the $p$-state becomes occupied and transitions between higher-lying states are possible. As seen in Figs.\ \ref{FigFIRpos}(b) and \ref{FigKap+transm}(c), then the upper mode, $\omega_+$, splits up into three resonances. This complex behavior of $\omega_+$ can be observed for $n_e = 3$, 4 and 5. For $n_e = 6$, the highest electron number that we can controllably load into the present dots, we again observe just two modes, however with a magnetic field dispersion that strongly deviates from that of a parabolic dot\cite{Fricke2}. A simple explanation for the occurrence of new resonances is given in the insets in Fig.\ \ref{FigFIRpos}: Due to a flattening out of the confining potential at higher energies, transitions with a smaller energy become possible when the $p$-state is partially occupied. Indeed, only these resonances remain observable when the $p$-state is completely filled and $s \rightarrow p$ transitions are no longer possible\cite{Fricke1}. The validity of such a single-particle explanation, however, is limited and cannot explain, e.g. the occurrence of a third $\omega_+$-mode. A theoretical model by Wojs and Hawrylak \cite{Hawrylak}, which takes into account many particle effects correctly predicts up to three $\omega_+$-modes when the $p$-state becomes partially filled.

From the solid lines in Fig.\ \ref{FigFIRpos}(a) we derive an effective mass of $m^* = 0.08$ $m_e$, significantly higher than that obtained by capacitance spectroscopy, above. A possible explanation for this is a magnetic field-dependent Coulomb contribution to the capacitance data (see below).

\begin{figure}
\vspace{5.5 cm}
\caption{(a) Relative splitting between the $s$-state charging peaks as a function of magnetic field, applied perpendicular (full symbols) and parallel (open symbols) to the plane of the dots. (b) Experimental and theoretical increase of the $s$-state splitting caused by a magnetic field-induced compression of the wave function.}
\label{FigSqueeze}
\end{figure}

As seen in Fig.\ \ref{FigBarbara}, the splitting between the $s$-state charging peaks is almost independent of magnetic field. Careful evaluation of the peak structure, however, shows that the energy $E_{1,2}$ increases by $\approx 2$\% when a magnetic field of 12 T is applied perpendicular to the plane of the dots (solid data points in Fig.\ \ref{FigSqueeze}(a)). This shift can be attributed in part to an increase of the Coulomb energy caused by a magnetic field-induced compression of the wave function. For parabolic confinement, the characteristic length of the ground state wave function $\ell=\sqrt{\hbar/(m^* \omega)}$ is magnetic field dependent through $\omega = \sqrt{\omega_0 + \omega_c^2 / 4}$, where $\omega_0 = \omega(B=0)$ and the cyclotron frequency $\omega_c$ are obtained from far-infrared spectroscopy. Similar to the classical Coulomb blockade, the energetic difference between the 1- and 2-electron ground state is inversely proportional to $\ell$, $E_{1,2}^{e-e} = e^2/(4\pi \varepsilon\varepsilon_0 \ell)$ \cite{MerktundCo}, which leads to an increased Coulomb blockade in high magnetic fields. 

Additionally, spin splitting will slightly increase $E_{1,2}$. The different contributions can be distinguished by their dependence on the direction of $B$: Because of the large confinement in the growth direction, only the spin splitting contributes to $E_{1,2}$ for parallel magnetic fields and the associated shift $E_{1,2}^{\parallel}$ (open symbols in Fig.\ \ref{FigSqueeze}(a)) is only about half of that for perpendicular fields, $E_{1,2}^{\perp}$. From $E_{1,2}^{\parallel}$ we can deduce an effective $g$-factor of $0.43 \pm 0.05$, which, as the effective mass, is in much better agreement with the surrounding GaAs than with the InAs of the dots.

The difference between $E_{1,2}^{\parallel}$ and $E_{1,2}^{\perp}$ is shown in Fig.\ \ref{FigSqueeze}(b). We attribute this difference to the magnetic field-induced compression of the ground state wave function, calculated using the above formula (solid line in Fig.\ \ref{FigSqueeze}(b)). The agreement between $\ell=4.4$ nm obtained from far-infrared spectroscopy and $\ell=4.9$ nm from capacitance, together with the agreement found in Fig.\ \ref{FigSqueeze}(b)) nicely demonstrates the compatibility of the different measurement techniques and the applicability of the models used.

\section{Acknowledgments}
Part of this work (GM-R and PMP) was supported through QUEST, a NSF Science and Technology Center. AL, BTM, and JPK acknowledge financial support through the DFG and the BMBF. The collaboration between Munich and Santa Barbara was supported through by a EC--US grant and a Max-Planck research award.


\begin{thebibliography}{44}
\bibitem{minute_splitting}
Here, we neglect a small ($\approx 2$ meV) splitting of the resonance at $B = 0$ due to a breaking of the symmetry caused either by growth anisotropy or by piezoelectric effects\protect\cite{Fricke1}.
\end{thebibliography}
\end{document}